\documentstyle[12pt]{article}

\makeatletter
\renewcommand\thesubsection{\thesection.\@arabic\c@subsection}
\makeatother

\newcommand{\sect}[1]{\setcounter{equation}{0}\section{#1}}

\newcommand {\beq}{\begin{equation}}
\newcommand {\eeq}{\end{equation}}
\newcommand {\beqa}{\begin{eqnarray}}
\newcommand {\eeqa}{\end{eqnarray}}         
\newcommand {\beqs}{\begin{eqnarray*}}
\newcommand {\eeqs}{\end{eqnarray*}}
\newcommand {\bds}{\begin{displaymath}}
\newcommand {\eds}{\end{displaymath}}
\newcommand {\n}{\nonumber\\}

\newcommand {\bebb}{\begin{thebibliography}}
\newcommand {\eebb}{\end{thebibliography}}      
\newcommand {\bbit}{\bibitem}

\def\pd{\prod}

\def\ra{\rangle}

\def\dg{\dagger}

\def\journal#1&#2(#3){\unskip, \sl #1\ \bf #2 \rm(19#3) }
\def\andjournal#1&#2(#3){\sl #1~\bf #2 \rm (19#3) }

\def\dz{\frac{d}{dz}}

\begin{document}


\begin{flushright}
\end{flushright}

\vskip 1cm

\begin{center}
{\Large\bf On the solvability of the quantum Rabi model and its 2-photon and two-mode generalizations}

\vspace{1cm}

{\large Yao-Zhong Zhang}
\vskip.1in

{\em School of Mathematics and Physics,
The University of Queensland, Brisbane, Qld 4072, Australia}

\end{center}

\date{}



\begin{abstract}
We study the solvability of the time-independent matrix Schr\"odinger differential equations of 
the quantum Rabi model and its 2-photon and two-mode generalizations in Bargmann Hilbert spaces of entire functions. 
We show that the Rabi model and its 2-photon and two-mode analogs are quasi-exactly solvable. 
We derive the exact, closed-form expressions for the energies and the allowed model parameters 
for all the three cases in the solvable subspaces. Up to a normalization factor, the eigenfunctions 
for these models are given by polynomials whose roots are determined by systems of algebraic equations.   
\end{abstract}

\vskip.1in





\setcounter{section}{0}
\setcounter{equation}{0}
\sect{Introduction}

The quantum Rabi model describes the interaction of a two-level atom with a single harmonic mode
of electromagnetic field. It is perhaps the simplest system for modeling the ubiquitous matter-light
interactions in modern physics, and has applications in a variety of physical fields, including
quantum optics \cite{Vedral06}, cavity and circuit quantum electrodynamics \cite{Englund07,Niemczyk10}, 
solid state semiconductor systems \cite{Khitrova06} and trapped ions \cite{Leibfried03}.

Recently Braak \cite{Braak11} presented a transcendental function defined as an infinite 
power series with coefficients satisfying a three-term recursive relation, 
and argued that the spectrum of the Rabi model
is given by the zeros of the transcendental function. 
This theoretical progress has renewed the interest in the Rabi and related models
\cite{Solano11,Moroz12,Maciejewski12,Travenec12,Chen12,Moroz13}.
However, 
since Braak's transcendental function is given as an infinite power series, 
unless the model parameters satisfy certain constraints for which the infinite
series truncates, its exact zeros and therefore  closed-form expressions for the energies
of the Rabi model can not be obtained even for those corresponding to the low-lying spectrum.

This is not surprising because, as pointed out in \cite{Moroz12,Moroz13}, 
the Rabi model is not exactly solvable but quasi-exactly solvable 
\cite{Turbiner88,Ushveridze94,Gonzarez93}. A typical feature of a quasi-exactly
solvable system is that only a finite part of the spectrum can be obtained
 in closed form and the remaining part of the spectrum is not algebraically accessible 
(i.e. can only be determined by numerical means).

In this paper we examine the solvability of the quantum Rabi model and its multi-mode 
and multi-quantum generalizations within the framework of Bargmann Hilbert spaces 
of entire functions \cite{Schweber67}. 
We show that the Rabi model and its  2-photon and two-mode generalizations are quasi-exactly solvable.  
We derive the explicit, closed-form expressions for the energies, the
allowed model parameters as well as the wavefunctions for all the three cases once and for all in terms
of the roots of the systems of algebraic equations. 
We note that the energies for the Rabi and 2-photon Rabi models were obtained previously 
\cite{Reik82,Kus85,Ng99,Emary02} by different methods.

The work is organized as follows. In section \ref{exact-quasi-exact} we recall the widely accepted
characterization of solvability of a linear differential operator in terms of invariants subspaces.
In sections \ref{rabi}, \ref{2-photon} and \ref{2-mode}, we obtain the exact solutions of 
the Rabi model and its 2-photon and 2-mode generalizations in their respective solvable subspaces.



\sect{Exact solvability versus quasi-exact solvability}\label{exact-quasi-exact}

Exact solvability and quasi-exact solvability are closely related \cite{Sasaki07}.
A quantum mechanical system is exactly solvable in the Schr\"odinger picture if all the
eigenvalues and the corresponding eigenfunctions of the system can be determined exactly.
In contrast, a system is quasi-exactly solvable 
if only a finite number of exact eigenvalues and eigenfunctions can be obtained. 
Among various characterizations of solvability, the one about the existence of invariant polynomial 
subspaces is conceptually the simplest. 
 
A linear differential operator ${H}$ is called quasi-exactly solvable if it has a finite-dimensional
invariant subspace ${\cal V}_{\cal N}$ with explicitly-described basis, that is,
$$
{H}{\cal V}_{\cal N}\subset {\cal V}_{\cal N},~~~{\rm dim}{\cal V}_{\cal N}<\infty, ~~~
{\cal V}_{\cal N}={\rm span}\{\xi_1,\cdots,\xi_{{\rm dim}{\cal V}_{\cal N}}\}.
$$
An immediate consequence of this characterization of quasi-exact solvablility for the operator $H$ 
is that it can be diagonalized {\em algebraically}
and exact, closed-form expressions of the corresponding spectra can be obtained
in the (solvable) subspace ${\cal V}_{\cal N}$. The remaining part of the spectrum is not 
analytically accessible 
and can only be computed through approximations (though sometimes rather accurately).
If the space ${\cal V}_{\cal N}$ 
is a subspace of a Bargmann-Hilbert space of entire functions in which $H$ is naturally defined,  
the solvable spectra and the corresponding vectors in ${\cal V}_{\cal N}$ give the exact 
eigenvalues and eigenfunctions of $H$, respectively.

A linear differential operator ${H}$ is exactly solvable if it preserves an infinite flag of 
finite-dimensional functional spaces, 
$$
{\cal V}_1\subset {\cal V}_2\subset\cdots \subset {\cal V}_{\cal N}\subset\cdots,
$$
whose bases admit explicit analytic forms, that is there exists a sequence of finite-dimensional 
invariant subspaces ${\cal V}_{\cal N},~{\cal N}=1,2,3,\cdots$,
$$
{H}{\cal V}_{\cal N}\subset {\cal V}_{\cal N},~~~{\rm dim}{\cal V}_{\cal N}<\infty, ~~~
{\cal V}_{\cal N}={\rm span}\{\xi_1,\cdots,\xi_{{\rm dim}{\cal V}_{\cal N}}\}.
$$

As will be seen in next sections, based on the above characterization of solvability of 
a quantum system, 
the Rabi model and its 2-photon and two-mode analogs are not exactly solvable, 
contrary to the claims by \cite{Braak11,Solano11,Travenec12,Chen12}. 
Rather they are quasi-exactly solvable, as was also noted by Moroz \cite{Moroz13} for the Rabi model,
because only a finite part of their spectra can be determined exactly.

\sect{Quasi-exact solvability of the quantum Rabi model}\label{rabi}
Quasi-exact solvability of the Rabi model has recently been noted by Moroz \cite{Moroz13}. 
Special exact spectrum of the model was obtained in \cite{Reik82,Kus85,Emary02} 
by different methods. In this section we re-examine this model within the framework of the
Bargmann Hilbert space of entire functions We will solve the
time-independent Schr\"odinger matrix differential equations by means of 
the functional Bethe ansatz method \cite{Lee10,Lee11,Zhang12}. In addition to the exact spectrum,
 we are also able to obtain the closed form
expressions for the allowed model parameters and the polynomial wavefunctions in terms of the roots
of a set of algebraic equations.
 
The Hamiltonian of the Rabi model is  
\begin{equation}
H=\omega a^\dagger a+\Delta\sigma_z+g\,\sigma_x\left[a^\dagger+a\right],\label{RabiH}
\end{equation}
where $g$ is the interaction strength, 
$\sigma_z, \sigma_x$ are the Pauli matrices describing the two atomic levels separated 
by energy difference $2\Delta$, and $a^\dagger$ ($a$) are 
creation (annihilation) operators of a boson mode with frequency $\omega$. 
In the Bargmann realization $a^\dagger\rightarrow z,~a\rightarrow\frac{d}{dz}$, 
the Hamiltonian becomes a matrix differential operator
\beq
H=\omega z\frac{d}{dz}+\Delta \sigma_z+g\,\sigma_x\left(z+\frac{d}{dz}\right).
\eeq
Working in a representation defined by $\sigma_x$ diagonal and in terms of the two-component wavefunction 
$\psi(z)=\left(\begin{array}{c}
\psi_+(z)\\
\psi_-(z)
\end{array} \right)$,
the time-independent Schr\"odinger equation gives rise to a coupled system of 
two 1st-order differential equations \cite{Braak11} 
\beqa
&&(\omega z+g)\frac{d}{dz}\psi_+(z)+(gz-E)\psi_+(z)+\Delta\psi_-(z)=0,\n
&&(\omega z-g)\frac{d}{dz}\psi_-(z)-(gz+E)\psi_-(z)+\Delta\psi_+(z)=0.
\eeqa
If $\Delta=0$ these two equations decouple and reduce to the differential equations of two uncoupled
displaced harmonic oscillators which can be exactly solved separately \cite{Zhang13a}. 
For this reason in the following we will concentrate on the $\Delta\neq 0$ case.

With the substitution $\psi_\pm(z)=e^{-gz/\omega}\phi_\pm(z)$, it follows \cite{Braak11}
\beqa
&&\left[(\omega z+g)\frac{d}{dz}-\left(\frac{g^2}{\omega}+E\right)\right]\phi_+(z)=-\Delta\phi_-(z),\n
&&\left[(\omega z-g)\frac{d}{dz}-\left(2gz-\frac{g^2}{\omega}+E\right)\right]\phi_-(z)=-\Delta\phi_+(z).
  \label{Rabi-diff}
\eeqa
The differential operator 
\beq
{\cal L}^1_R\equiv (\omega z+g)\frac{d}{dz}-\left(\frac{g^2}{\omega}+E\right)
\eeq
in the first equation is then exactly solvable.
Eliminating $\phi_-(z)$ from the system we obtain the uncoupled differential equation for $\phi_+(z)$,
\beqa
\left[(\omega z-g)\frac{d}{dz}-\left(2gz-\frac{g^2}{\omega}+E\right)\right]
  \left[(\omega z+g)\frac{d}{dz}-\left(\frac{g^2}{\omega}+E\right)\right]\phi_+=\Delta^2\phi_+.
\eeqa
This is a second-order differential equation of Fuchs' type. Explicitly, 
\beqa
&&(\omega z-g)(\omega z+g)\frac{d^2\phi_+}{dz^2}+\left[-2\omega gz^2+(\omega^2-2g^2-2E\omega)z
   +\frac{g}{\omega}(2g^2-\omega^2)\right]\frac{d\phi_+}{dz}\n
&&~~~~~~~~~~~~~~~~~~+\left[2g\left(\frac{g^2}{\omega}+E\right)z+E^2-\Delta^2-\frac{g^4}{\omega^2}
   \right]\phi_+\equiv {\cal L}\phi_+=0.\label{Rabi-diff1}
\eeqa
It is easy to see that for any positive integer $n$,
\beqa
{\cal L}z^n&=&\left[-2n\omega g+2g\left(E+\frac{g^2}{\omega}\right)\right]z^{n+1}\n
& &+   \left[n(n-1)\omega^2+(\omega^2-2g^2-2\omega E)n+E^2-\Delta^2-\frac{g^4}{\omega^2}\right]z^n\n
& &   +{\rm lower~order~terms}.\label{Rabi-quasi-exact-solvability-test}
\eeqa
Due to the $z^{n+1}$ term on the right hand side of (\ref{Rabi-quasi-exact-solvability-test}), 
the operator ${\cal L}$ is not exactly solvable. Only if $n={\cal N}$ such that 
$-2{\cal N}\omega g+2g\left(E+\frac{g^2}{\omega}\right)=0$, the $z^{n+1}$ term disappears 
from the right hand side, and ${\cal L}$ preserves a finite dimensional subspace 
${\cal V}_{\cal N}={\rm span}\{1,z, z^2,\cdots,z^{\cal N}\}$. Therefore ${\cal L}$ is quasi-exactly
solvable with invariant subspace ${\cal V}_{\cal N}={\rm span}\{1, z, z^2,\cdots, z^{\cal N}\}$.

We now seek exact solutions of the differential equation in the solvable subspace ${\cal V}_{\cal N}$.
Obviously they are polynomials in $z$ of degree ${\cal N}$, which can be written as
\beqa
\phi_+(z)&=&\pd_{i=1}^{\cal N}(z-z_i),~~~~{\cal N}=1,2,\cdots,
\eeqa
where $z_i$ are the roots of the polynomial to be determined. The energies follow immediately
from the equation in the 2nd line below (\ref{Rabi-quasi-exact-solvability-test}),
\beq
E=\omega\left({\cal N}-\frac{g^2}{\omega^2}\right).\label{Rabi-solution-E}
\eeq
The differential equation (\ref{Rabi-diff1}) is in fact a special case of the general equation 
studied in \cite{Zhang12}.  Applying the results of this reference, we obtain
the constraint for the system parameters
\beq
\Delta^2+2{\cal N}g^2+2\omega g\sum_{i=1}^{\cal N}z_i=0.\label{Rabi-solution-constraint}
\eeq
Here $z_i$ satisfy the set of algebraic equations
\beq
\sum_{j\neq i}^{\cal N}\frac{2}{z_i-z_j}=\frac{2\omega g z_i^2+(2{\cal N}-1)\omega^2 z_i
  +g(\omega^2-2g^2)/\omega}{(\omega z_i-g)(\omega z_i+g)},~~~~i=1,2,\cdots, {\cal N}.
  \label{Rabi-solution-BEs}
\eeq
The corresponding wavefunction component $\psi_+(z)$ of the model is then given by
\beq
\psi_+(z)=e^{-\frac{g}{\omega}z}\prod_{i=1}^{\cal N}(z-z_i),
\eeq
and the component $\psi_-(z)=e^{-gz/\omega}\phi_-(z)$ with $\phi_-(z)$ determined by 
the first equation of (\ref{Rabi-diff}) for $\Delta\neq 0$, i.e. $\phi_-(z)=-\frac{1}{\Delta}{\cal L}^1_R\,\phi_+(z)$.
Because ${\cal L}^1_R$ preserves ${\cal V}_{\cal N}$ for any system parameters, 
$\phi_-(z)$ automatically belongs to the same invariant subspace as $\phi_+(z)$. 

As an example to the above general expressions, let us consider the ${\cal N}=1$ solution.
The energy is $E=\omega\left(1-\frac{g^2}{\omega^2}\right)$. 
(\ref{Rabi-solution-BEs}) becomes 
\beq
2\omega gz_1^2+\omega^2 z_1+\frac{g}{\omega}(\omega^2-2g^2)=0,
\eeq
which has two solutions 
\beq
z_1=-\frac{g}{\omega},~~~~~~\frac{2g^2-\omega^2}{2\omega g}.
\eeq
Substituting into (\ref{Rabi-solution-constraint}) gives the constraints $\Delta=0$ and 
$\Delta^2+4g^2=\omega^2$, respectively.  The constraint $\Delta=0$ corresponds to the case of
degenerate atomic levels. The other constraint agrees with that obtained in \cite{Emary02} by
a different approach. The corresponding wave function is 
\beq
\psi_+(z)=e^{-\frac{g}{\omega}z}\left(z-\frac{2g^2-\omega^2}{2\omega g}\right).
\eeq

\sect{Quasi-exact solvability of the 2-photon quantum Rabi model}\label{2-photon}
The Hamiltonian of the 2-photon Rabi model reads
\begin{equation}
H=\omega a^\dagger a+\Delta\sigma_z+g\,\sigma_x\left[(a^\dagger)^2+a^2\right].\label{2-photon-RabiH1}
\end{equation}
Introduce the operators $K_\pm, K_0$ by \cite{Ng99,Emary02,Gerry85}
\beq
K_+=\frac{1}{2}(a^\dagger)^2,~~~~K_-=\frac{1}{2}a^2,~~~~K_0=\frac{1}{2}
\left(a^\dagger a+\frac{1}{2}\right).   \label{2-photon-Rabi-boson}
\eeq
Then the Hamiltonian (\ref{2-photon-RabiH1}) becomes \cite{Ng99,Emary02}
\beq
H=2\omega\left(K_0-\frac{1}{4}\right)+\Delta\sigma_z+2g\,\sigma_x(K_++K_-).\label{2-photon-RabiH2}
\eeq
The operators $K_\pm, K_0$ form the usual $su(1,1)$ Lie algebra,
\beq
[K_0, K_\pm]=\pm K_\pm,~~~~~ [K_+, K_-]=-2K_0.\label{su11-relations}
\eeq
The quadratic Casimir operator $C$ of the algebra is given by 
\beq
C=K_+K_--K_0(K_0-1).\label{su11-casimir}
\eeq
In what follows we shall use an infinite-dimensional unitary irreducible representation of 
$su(1,1)$ known as the positive discrete series ${\cal D}^+(q)$. The parameter $q$ is the 
so-called Bargmann index. In this representation the basis states $\{|q,n\ra\}$ diagonalize the
operator $K_0$,
\beq
K_0|q,n\ra=(n+q)|q,n\ra\label{K0-rep}
\eeq
for $q>0$ and $n=0,1,2, \cdots$, and the Casimir operator $C$ has the eigenvalue $q(1-q)$.
The operators $K_+$ and $K_-$ are hermitian to each other 
and act as raising and lowering operators respectively within ${\cal D}^+(q)$, 
\beqa
K_+|q,n\ra&=&\sqrt{(n+1)(n+2q)}\;|q,n+1\ra,\n
K_-|q,n\ra&=&\sqrt{n(n+2q-1)}\;|q,n-1\ra.\label{su11-rep}
\eeqa

For the single-mode bosonic realization (\ref{2-photon-Rabi-boson}), $C=\frac{3}{16}$ and 
$q$ is equal to either $\frac{1}{4}$ or $\frac{3}{4}$. 
In terms of the original Bose operators the states $|q,n\ra$ are given equivalently as
\beq
|q,n\ra=\frac{(a^\dagger)^{2(n+q-\frac{1}{4})}}{\sqrt{\left[2\left(n+q-\frac{1}{4}\right)\right]!}}|0\ra,~~~~q=1/4,~3/4;~~~~n=0,1,2,\cdots.
\eeq
Thus by means of the $su(1,1)$ representation, we have decomposed the Fock-Hilbert 
space of the boson field into the direct sum of two independent subspaces labeled by 
$q=1/4, 3/4$, respectively. 

Let us now derive a single variable differential operator realization of $K_\pm, K_0$
(\ref{2-photon-Rabi-boson}), i.e. differential realization of the infinite-dimensional unitary
 irreducible representation  corresponding to $q=1/4, 3/4$.
Using the Fock-Bargmann correspondence  
$a^{\dg} \rightarrow w,~~ a \rightarrow \frac{d}{dw},~~ |0\ra \rightarrow 1$,
we can make the association
$$
|q,n\ra\longrightarrow \frac{w^{2(n+q-1/4)}}{\sqrt{\left[ 2\left(n+q-\frac{1}{4}\right)\right]!}}
=\frac{w^{2q-1/2}\,(w^2)^n}{\sqrt{\left[ 2\left(n+q-\frac{1}{4}\right)\right]!}}.
$$
Since $q$ is constant for a given representation we can rewrite the above as a mapping
of the Fock states $|q,n\ra$ to the monomials in $z=w^2$, 
\beqa
\Psi_{q,n}(z)=\frac{z^n}{\sqrt{\left[ 2\left(n+q-\frac{1}{4}\right)\right]!}},~~~~q=1/4,~3/4;~~~~
 n=0,1,2,\cdots.\label{2-photon-monimials}
\eeqa
Then in the Bargmann space with basis vectors $\Psi_{q,n}(z)$, 
the operators $K_\pm, K_0$ (\ref{2-photon-Rabi-boson}) are realized by 
single-variable 2nd differential operators as
\beq
{K}_0 = z\frac{d}{dz}+ q, ~~~~{K}_+ = \frac{z}{2}, ~~~~
{K}_- =2z\frac{d^2}{dz^2}+4q\frac{d}{dz}.\label{2-photon-Rabi-diff}
\eeq
It can be checked that the differential operators (\ref{2-photon-Rabi-diff}) satisfy the $su(1,1)$ 
commutation relations (\ref{su11-relations}) and their action on $\Psi_{q,n}(z)$ gives the 
representation (\ref{K0-rep}, \ref{su11-rep}) corresponding to $q=1/4, 3/4$.  In checking e.g.
$K_-\Psi_{q,n}(z)=\sqrt{n(n+2q-1)}\;\Psi_{q,n-1}(z)$, it is useful to note that 
$(q-1/4)(q-3/4)\equiv 0$ for both $q=1/4, 3/4$ and the differential
operator $K_-$ above can be expressed as
$K_- =2z^{-1}\left(z\frac{d}{dz}+q-\frac{1}{4}\right)\left(z\frac{d}{dz}+q-\frac{3}{4}\right)$. 

In terms of the differential realization (\ref{2-photon-Rabi-diff}), the 2-photon Rabi Hamiltonian becomes
\beq
H=2\omega\left(z\frac{d}{dz}+q-\frac{1}{4}\right)+\Delta\sigma_z+2g\,\sigma_x\left(\frac{z}{2}+2z\frac{d^2}{dz^2}+4q\frac{d}{dz}\right).
\eeq
Working in a representation defined by $\sigma_x$ diagonal and in terms of the two component wavefunction,
the time-independent Schr\"odinger equation leads to two coupled 2nd-order differential equations,
\beqa
&&4gz\frac{d^2}{dz^2}\psi_+(z)+(2\omega z+8gq)\frac{d}{dz}\psi_+(z)
   +\left[gz+2\omega\left(q-\frac{1}{4}\right)-E\right]\psi_+(z)+\Delta\psi_-(z)=0,\n
&&4gz\frac{d^2}{dz^2}\psi_-(z)+(-2\omega z+8gq)\frac{d}{dz}\psi_-(z)
   +\left[gz-2\omega\left(q-\frac{1}{4}\right)+E\right]\psi_-(z)-\Delta\psi_+(z)=0.\n
\eeqa
If $\Delta=0$ these two equations decouple and reduce to the differential equations of two uncoupled
single-mode squeezed harmonic oscillators which can be exactly solved separately \cite{Zhang13a}. 
For this reason in the following we will concentrate on the $\Delta\neq 0$ case.

With the substitution
\beq
\psi_\pm(z)=e^{-\frac{\omega}{4g}  (1-\Omega) z}\varphi_\pm(z),~~~~~~\Omega=\sqrt{1-\frac{4g^2}{\omega^2}},\label{2-photon-sustitution}
\eeq
where $\left|\frac{2g}{\omega}\right|<1$, it follows,
\beqa
&&\left\{4gz\frac{d^2}{dz^2}+[2\omega\Omega z+8gq]\frac{d}{dz}
  +2q\omega\Omega-\frac{1}{2}\omega-E\right\}\varphi_+=-\Delta \varphi_-,\n
&&\left\{4gz\frac{d^2}{dz^2}+[2\omega(\Omega-2)z+8gq]\frac{d}{dz}
  +\frac{\omega^2}{g}(1-\Omega)z+ 2q\omega(\Omega-2)+\frac{1}{2}\omega+E\right\}\varphi_-=\Delta \varphi_+.\n
  \label{2-photon-diff}
\eeqa
Then the  differential operator
\beq
{\cal L}_{2-p}\equiv 4gz\frac{d^2}{dz^2}+[2\omega\Omega z+8gq]\frac{d}{dz}
  +2q\omega\Omega-\frac{1}{2}\omega-E
\eeq
appearing in the first equation is exactly solvable.
Eliminating $\varphi_-(z)$ from the system, we obtain the uncoupled differential equation for $\varphi_+(z)$
\beqa
&&\left\{4gz\frac{d^2}{dz^2}+[2\omega(\Omega-2)z+8gq]\frac{d}{dz}
   +\frac{\omega^2}{g}(1-\Omega)z+ 2q\omega(\Omega-2)+\frac{1}{2}\omega+E\right\}\n
&&~~~\times \left\{4gz\frac{d^2}{dz^2}+[2\omega\Omega z+8gq]\frac{d}{dz}+2q\omega\Omega-\frac{1}{2}\omega-E\right\}\varphi_+(z)=-\Delta^2 \varphi_+(z). 
\eeqa
This is a 4th-order differential equation of Fuchs' type. Explicitly,
\beqa
&&16g^2z^2\frac{d^4\varphi_+}{dz^4}+64g^2\left[\frac{\omega}{4g}(\Omega-1) z^2+\left(q+\frac{1}{2}\right)z\right]\frac{d^3\varphi_+}{dz^3}\n
&&~~~~+\left\{4\omega^2(\Omega^2-3\Omega+1)z^2+16\omega g\left[3\left(q+\frac{1}{2}\right)\Omega-3q-1\right]z+64g^2q\left(q+\frac{1}{2}\right)\right\}\frac{d^2\varphi_+}{dz^2}\n
&&~~~~+\left\{2\frac{\omega^3}{g}\Omega(1-\Omega)z^2+\left[8\omega^2q(1-\Omega)+8\omega^2\left(q+\frac{1}{2}\right)(1-\Omega)^2\right.\right.\n
&&~~~~~~~~~~\left.\left.     +4\omega\left(E-2\omega\left(q+\frac{1}{4}\right)\right)\right]z  +32\omega gq\left[\left(q+\frac{1}{2}\right)\Omega-q\right]\right\}\frac{d\varphi_+}{dz}\n
&&~~~~+\left\{\frac{\omega^2}{g}(1-\Omega)\left(2q\omega\Omega-\frac{1}{2}\omega-E\right)z\right.\n
&&~~~~~~~~~~\left. +4\omega^2q^2(1-\Omega)^2-\left[E-2\omega\left(q-\frac{1}{4}\right)\right]^2+\Delta^2\right\}\varphi_+=0.\label{2-photon-diff1}
\eeqa
By the arguments similar to those in the previous section, 
this equation is quasi-exactly solvable provided that the system parameters 
$\Delta, \omega$ and $g$ satisfy certain constraints, and exact solutions are polynomials in $z$ in the 
solvable sector. We thus seek polynomial solutions of the form to the above differential equation, 
\beqa
\varphi_+(z)&=&\pd_{i=1}^{\cal M}(z-z_i),~~~~{\cal M}=1,2,\cdots,
\eeqa
where ${\cal M}$ is the degree of the polynomial solution and $z_i$ are the roots of the polynomial 
to be determined. Substituting into (\ref{2-photon-diff1}) and dividing both sides by $\phi_+(z)$ yield
\beqa
&&\left[E-2\omega\left(q-\frac{1}{4}\right)\right]^2-\Delta^2-4\omega^2q^2(1-\Omega)^2\n
&&~~~~~~~~~=16g^2z^2\sum_{i=1}^{\cal M}\frac{1}{z-z_i}
    \sum_{p\neq l\neq j\neq i}^{\cal M}\frac{4}{(z_i-z_p)(z_i-z_l)(z_i-z_j)}\n
&&~~~~~~~~~~~~+64g^2\left[\frac{\omega}{4g}(\Omega-1) z^2+\left(q+\frac{1}{2}\right)z\right]
    \sum_{i=1}^{\cal M}\frac{1}{z-z_i}
    \sum_{l\neq j\neq i}^{\cal M}\frac{3}{(z_i-z_l)(z_i-z_j)}\n
&&~~~~~~~~~~~~+\left\{4\omega^2(\Omega^2-3\Omega+1)z^2+16\omega g\left[3\left(q+\frac{1}{2}\right)\Omega-3q-1\right]z\right.\n
&&~~~~~~~~~~~~~~~~~~~\left. +64g^2q\left(q+\frac{1}{2}\right)\right\}\sum_{i=1}^{\cal M}\frac{1}{z-z_i}
    \sum_{j\neq i}^{\cal M}\frac{2}{z_i-z_j}\n
&&~~~~~~~~~~~~+\left\{2\frac{\omega^3}{g}\Omega(1-\Omega)z^2+\left[8\omega^2q(1-\Omega)+8\omega^2\left(q+\frac{1}{2}\right)(1-\Omega)^2\right.\right.\n
&&~~~~~~~~~~~~~~~~~~~\left.\left. +4\omega\left(E-2\omega\left(q+\frac{1}{4}\right)\right)\right]z  
    +32\omega gq\left[\left(q+\frac{1}{2}\right)\Omega-q\right]\right\}
    \sum_{i=1}^{\cal M}\frac{1}{z-z_i}\n
&&~~~~~~~~~~~~+ \frac{\omega^2}{g}(1-\Omega)\left(2q\omega\Omega-\frac{1}{2}\omega-E\right)z.
\eeqa
The left hand side is a constant and the right hand side is a meromorphic function 
with simples poles at $z=z_i$ and
singularity at $z=\infty$. The residues of the right hand side at the simple poles $z=z_i$ are
\beqa
{\rm Res}_{z=z_i}&=&16g^2z_i^2\sum_{p\neq l\neq j\neq i}^{\cal M}\frac{4}{(z_i-z_p)(z_i-z_l)(z_i-z_j)}\n
& &+64g^2\left[\frac{\omega}{4g}(\Omega-1) z_i^2+\left(q+\frac{1}{2}\right)z_i\right]
   \sum_{l\neq j\neq i}^{\cal M}\frac{3}{(z_i-z_l)(z_i-z_j)}\n
& &+\left\{4\omega^2(\Omega^2-3\Omega+1)z_i^2+16\omega g\left[3\left(q+\frac{1}{2}\right)
    \Omega-3q-1\right]z_i\right.\n
& &~~~~~\left. +64g^2q\left(q+\frac{1}{2}\right)\right\}\sum_{j\neq i}^{\cal M}\frac{2}{z_i-z_j}\n
& &+2\frac{\omega^3}{g}\Omega(1-\Omega)z_i^2+\left[8\omega^2 q(1-\Omega)
    +8\omega^2\left(q+\frac{1}{2}\right)(1-\Omega)^2\right.\n
& &~~~~~\left. +4\omega\left(E-2\omega\left(q+\frac{1}{4}\right)\right)\right]z_i  
    +32\omega gq\left[\left(q+\frac{1}{2}\right)\Omega-q\right].\label{2-photon-residues}
\eeqa
Using the identities
\beqa
&&\sum_{i=1}^{\cal M}\sum_{j\neq i}^{\cal M}\frac{1}{z_i-z_j}=0,~~~~~
  \sum_{i=1}^{\cal M}\sum_{j\neq i}^{\cal M}\frac{z_i}{z_i-z_j}=\frac{1}{2}{\cal M}({\cal M}-1),\n
&&\sum_{i=1}^{\cal M}\sum_{l\neq j\neq i}^{\cal M}\frac{1}{(z_i-z_l)(z_i-z_j)}=0,~~~~~
  \sum_{i=1}^{\cal M}\sum_{l\neq j\neq i}^{\cal M}\frac{z_i}{(z_i-z_l)(z_i-z_j)}=0,\n
&&\sum_{i=1}^{\cal M}\sum_{p\neq l\neq j\neq i}^{\cal M}\frac{1}{(z_i-z_p)(z_i-z_l)(z_i-z_j)}=0,\n
&&\sum_{i=1}^{\cal M}\sum_{p\neq l\neq j\neq i}^{\cal M}\frac{z_i}{(z_i-z_p)(z_i-z_l)(z_i-z_j)}=0,
\eeqa
we can show that
\beqa
&&\left[E-2\omega\left(q-\frac{1}{4}\right)\right]^2-\Delta^2-4\omega^2q^2(1-\Omega)^2\n
&&~~~~~~~~~~~~~= \sum_{i=1}^{\cal M}\frac{{\rm Res}_{z=z_i}}{z-z_i}
   +4\omega^2(\Omega^2-3\Omega+1){\cal M}({\cal M}-1)
   +2\frac{\omega^3}{g}\Omega(1-\Omega)\sum_{i=1}^{\cal M}z_i\n
&&~~~~~~~~~~~~~~~+\left[8\omega^2 q(1-\Omega)
    +8\omega^2\left(q+\frac{1}{2}\right)(1-\Omega)^2+4\omega\left(E-2\omega\left(q+\frac{1}{4}\right)\right)\right]{\cal M}\n
&&~~~~~~~~~~~~~~~+\frac{\omega^2}{g}(1-\Omega)\left[(2{\cal M}+2q)\omega\Omega
    -\frac{1}{2}\omega-E\right]z.\label{2-photon-mid-steps}
\eeqa
The right hand side of (\ref{2-photon-mid-steps}) is a constant if and only if the coefficient of
$z$ as well as all the residues at the simple poles are vanishing. We thus obtain the energy eigenvalues,
\beqa
&&E=-\frac{1}{2}\omega+\left[2{\cal M}+2\left(q-\frac{1}{4}\right)+\frac{1}{2}\right]\omega\Omega,\label{2-photon-energy}
\eeqa
and the constraint for the model parameters $\Delta, \omega$ and $g$, 
\beqa
\Delta^2+4\omega^2(1-\Omega)\left[{\cal M}({\cal M}+2q-1)+\frac{\omega}{2g}\Omega
   \sum_{i=1}^{\cal M}z_i\right]=0.   \label{2-photon-constraint}
\eeqa
Here the roots $z_i$ satisfy the following system of algebraic equations,
\beqa
&&g^2z_i^2\sum_{p\neq l\neq j\neq i}^{\cal M}\frac{4}{(z_i-z_p)(z_i-z_l)(z_i-z_j)}\n
&&~~~~~+g\left[\omega(\Omega-1) z_i^2+4g\left(q+\frac{1}{2}\right)z_i\right]
   \sum_{l\neq j\neq i}^{\cal M}\frac{3}{(z_i-z_l)(z_i-z_j)}\n
&&~~~~~+\left\{\frac{\omega^2}{4}(\Omega^2-3\Omega+1)z_i^2+\omega g\left[3\left(q+\frac{1}{2}\right)
    \Omega-3q-1\right]z_i+4g^2q\left(q+\frac{1}{2}\right)\right\}\sum_{j\neq i}^{\cal M}\frac{2}{z_i-z_j}\n
&&~~~~~+\frac{\omega^3}{8g}\Omega(1-\Omega)z_i^2+\frac{\omega^2}{2}\left[{\cal M}\Omega+\left(q+\frac{1}{2}\right)\Omega(\Omega-2)+q\right]z_i\n
&&~~~~~+2\omega gq\left[\left(q+\frac{1}{2}\right)\Omega-q\right]~=~0,~~~~~~~~i=1,2,\cdots {\cal M}.\label{2-photon-BEs}
\eeqa
Here we have also used the relation (\ref{2-photon-energy}) in obtaining (\ref{2-photon-constraint}) 
and (\ref{2-photon-BEs}). The corresponding wavefunction component $\psi_+(z)$ of the system is
given by
\beq
\psi_+(z)=e^{-\frac{\omega}{4g}(1-\Omega)z}\prod_{i=1}^{\cal M}(z-z_i),
\eeq
and the component $\psi_-(z)=e^{-\frac{\omega}{4g}(1-\Omega)z}\phi_-(z)$ with $\phi_-(z)$
computed from (\ref{2-photon-diff}) for $\Delta\neq 0$, 
$\phi_-(z)=-\frac{1}{\Delta}{\cal L}_{2-p}\,\phi_+(z)$. Because for any positive integer $n$, 
${\cal L}_{2-p}z^n\sim z^n+{\rm lower~order~terms}$,
$\phi_-(z)$ automatically belongs to the same invariant subspace as $\phi_+(z)$.

Some remarks are in order. The results obtained in \cite{Dolya09} on the solution of 
the 2-photon Rabi model are
incorrect. The reason is simple. As the author himself noted in the paper, the solution for 
the 2nd component of the two-component wavefunction of the 2-photon Rabi model does not 
belong to the same
invariant subspace as the 1st component. Therefore his two-component wavefunction does not 
satisfy the matrix Schr\"odinger differential equation of the model and thus is not a solution
to the coupled equations.

As an example to the above general expressions, let us consider the ${\cal M}=1$ case. The energy is
\beq
E=-\frac{1}{2}\omega+\left(2q+2\right)\omega\Omega.
\eeq
(\ref{2-photon-BEs}) becomes 
\beqa
&&\omega^2\Omega(1-\Omega)z_1^2+4\omega g\left[\Omega+q+\left(q+\frac{1}{2}\right)\Omega(\Omega-2)\right]z_i
+16g^2q\left[\left(q+\frac{1}{2}\right)\Omega-q\right]~=~0.\n
\eeqa
The solutions to this equation are 
\beq
z_1=-\frac{4gq}{\omega\Omega},~~~~~~\frac{4gq(1-\Omega)-2g\Omega}{\omega(1-\Omega)}.
\eeq
Substituting into (\ref{2-photon-constraint}) gives the constraints $\Delta=0$ and
\beq
\Delta^2+8q\omega^2=8\left(q+\frac{1}{2}\right)\omega^2\Omega^2,\label{2-photon-constraint-example}
\eeq
respectively. The constraint $\Delta=0$ corresponds to the case of degenerate atomic levels. The
constraint (\ref{2-photon-constraint-example}) agrees with those obtained in \cite{Emary02} by
the Bogoliubov transformation method. The corresponding wavefunction $\psi_+(z)$ is given by
\beq
\psi_+(z)=e^{-\frac{\omega}{4g}(1-\Omega)z}\left(z-\frac{4gq(1-\Omega)-2g\Omega}{\omega(1-\Omega)}\right).
\eeq

\sect{Quasi-exact solvability of the two-mode quantum Rabi model}\label{2-mode}
The Hamiltonian of the two-mode quantum Rabi model reads
\beq
H=\omega(a_1^\dagger a_1+a_2^\dagger a_2)+\Delta\sigma_z+g\,\sigma_x(a_1^\dagger a_2^\dagger+a_1 a_2),
  \label{2-mode-RabiH1}
\eeq
where we assume that the boson modes are degenerate with the same frequency $\omega$.
Introduce the operators $K_\pm, K_0$,
\beq
K_+=a_1^\dagger a_2^\dagger,~~~~K_-=a_1 a_2,~~~~K_0=\frac{1}{2}(a_1^\dagger a_1+a_2^\dagger a_2+1).\label{2-mode-Rabi-boson}
\eeq
Then the Hamiltonian (\ref{2-mode-RabiH1}) becomes
\beq
H=2\omega\left(K_0-\frac{1}{2}\right)+\Delta\sigma_z+g\sigma_x(K_++K_-).\label{2-mode-RabiH2}
\eeq
The operators $K_\pm, K_0$ form the $su(1,1)$ algebra (\ref{su11-relations}). As in the previous 
section we shall use the unitary irreducible representation (i.e. the positive discrete series). 
However, to avoid confusion in this section we shall use $\kappa$ to denote the Bargmann 
index of the representation. Using this notation the action of the operators $K_\pm, K_0$ and 
the Casimir $C$ (\ref{su11-casimir}) on the basis states  $|\kappa,n\ra$ of the representation reads
\beqa
K_0|\kappa,n\ra&=&(n+\kappa)\;|\kappa,n\ra,\n
K_+|\kappa,n\ra&=&\sqrt{(n+2\kappa)(n+1)}\;|\kappa,n+1\ra,\n
K_-|\kappa,n\ra&=&\sqrt{(n+2\kappa-1)n}\;|\kappa,n-1\ra,\n
C|\kappa,n\ra&=&\kappa(1-\kappa)\;|\kappa,n\ra,\label{2-mode-rep}
\eeqa
for $\kappa>0$ and $n=0,1,2,\cdots$.

For the two-mode bosonic realization (\ref{2-mode-Rabi-boson}) of $su(1,1)$ that we require here
the Bargmann index $\kappa$ can take any positive integers or half-integers, 
i.e. $\kappa=1/2, 1, 3/2,\cdots$. In terms of the original Bose operators the states 
$|\kappa,n\ra$ are given equivalently as 
\beq
|\kappa,n\ra=\frac{(a_1^\dagger)^{n+2\kappa-1} (a_2^\dagger)^n}{\sqrt{(n+2\kappa-1)!n!}}|0\ra,~~~~~
    \kappa=1/2, 1, 3/2, \cdots;~~~~~n=0,1,2,\cdots.
\eeq
Thus by means of the $su(1,1)$ representation we have decomposed the Fock-Hilbert space 
into the direct sum of infinite number of subspaces labeled by 
$\kappa=1/2, 1, 3/2, \cdots$, respectively.

As in the previous section,  using the Fock-Bargmann correspondence
$a_i^{\dg} \rightarrow w_i, ~~ a_i \rightarrow \frac{d}{dw_i}, ~~ |0\ra \rightarrow 1$,
we can make the association
$$
|\kappa,n\ra\longrightarrow \frac{w_1^{n+2\kappa-1}w_2^n}{\sqrt{(n+2\kappa-1)!n!}}=
  \frac{w_1^{2\kappa-1}\, (w_1w_2)^n}{\sqrt{(n+2\kappa-1)!n!}}.
$$
Then, with $\kappa$ being constant in a given representation, we can
show that the set of monomials in $z=w_1w_2$,
\beq
\Psi_{\kappa,n}(z)=\frac{z^n}{\sqrt{(n+2\kappa-1)!n!}},~~~~\kappa=1/2, 1, 3/2,\cdots;~~~~n=0,1,2,\cdots,
\eeq
forms the basis carrying the unitary irreducible representation (\ref{2-mode-rep}) corresponding to
$\kappa=1/2, 1, 3/2,\cdots$. That is the operators $K_\pm, K_0$ 
(\ref{2-mode-Rabi-boson}) have the single-variable differential realization,
\beq
K_0=z\frac{d}{dz}+\kappa,~~~~K_+=z,~~~~K_-=z\frac{d^2}{dz^2}+2\kappa\frac{d}{dz}.
   \label{su11-diff-rep-2mode}
\eeq
It is straightforward to verify that these differential operators satisfy the $su(1,1)$ 
commutation relations (\ref{su11-relations}) and their action on $\Psi_{\kappa,n}(z)$ gives the unitary
representation (\ref{2-mode-rep}) corresponding to $\kappa=1/2, 1, 3/2,\cdots$.

By means of the differential representation (\ref{su11-diff-rep-2mode}), we can express the Hamiltonian 
(\ref{2-mode-RabiH2}) as the 2nd-order matrix differential operator
\beq
H=2\omega\left(z\frac{d}{dz}+\kappa-\frac{1}{2}\right)+\Delta\sigma_z+g\,\sigma_x\left(z+z\frac{d^2}{dz^2}
   +2\kappa\frac{d}{dz}\right).   \label{2-mode-Rabi-diff}
\eeq
Working in a representation defined by $\sigma_x$ diagonal and in terms of the two-component
wavefunction $\psi(z)=\left(\begin{array}{c}
\psi_+(z)\\
\psi_-(z)
\end{array} \right)$, we see that the time-independent Schr\"odinger equation 
yields the two coupled differential equations,
\beqa
&&gz\frac{d^2}{dz^2}\psi_+(z)+2(\omega z+g\kappa)\frac{d}{dz}\psi_+(z)
   +\left[gz+2\omega\left(\kappa-\frac{1}{2}\right)-E\right]\psi_+(z)+\Delta\psi_-(z)=0,\n
&&gz\frac{d^2}{dz^2}\psi_-(z)+2(-\omega z+g\kappa)\frac{d}{dz}\psi_-(z)
   +\left[gz-2\omega\left(\kappa-\frac{1}{2}\right)+E\right]\psi_-(z)-\Delta\psi_+(z)=0.\n
\eeqa
If $\Delta=0$ these two equations decouple and reduce to the differential equations of two uncoupled
two-mode squeezed harmonic oscillators which can be exactly solved separately \cite{Zhang13a}. 
For this reason in the following we will concentrate on the $\Delta\neq 0$ case.

With the substitution
\beq
\psi_\pm(z)=e^{-\frac{\omega}{g}  (1-\Lambda) z}\varphi_\pm(z),~~~~~~\Lambda=\sqrt{1-\frac{g^2}{\omega^2}},
\eeq
where $\left|\frac{g}{\omega}\right|<1$, it follows,
\beqa
&&\left\{gz\frac{d^2}{dz^2}+2[\omega\Lambda z+g\kappa]\frac{d}{dz}
  +2\kappa\omega\Lambda-\omega-E\right\}\varphi_+=-\Delta \varphi_-,\n
&&\left\{gz\frac{d^2}{dz^2}+2[\omega(\Lambda-2)z+g\kappa]\frac{d}{dz}
  +\frac{4\omega^2}{g}(1-\Lambda)z+ 2\kappa\omega(\Lambda-2)+\omega+E\right\}\varphi_-=\Delta \varphi_+.\n
  \label{2-mode-diff}
\eeqa
Then the differential operator
\beq
{\cal L}_{2-m}\equiv gz\frac{d^2}{dz^2}+2[\omega\Lambda z+g\kappa]\frac{d}{dz}  +2\kappa\omega\Lambda-\omega-E
\eeq
in the 1st equation is exactly solvable.
Eliminating $\varphi_-(z)$ from the system, we obtain the uncoupled differential equation for $\varphi_+(z)$
\beqa
&&\left\{gz\frac{d^2}{dz^2}+2[\omega(\Lambda-2)z+g\kappa]\frac{d}{dz}
   +\frac{4\omega^2}{g}(1-\Lambda)z+ 2\kappa\omega(\Lambda-2)+\omega+E\right\}\n
&&~~~\times \left\{gz\frac{d^2}{dz^2}+2[\omega\Lambda z+g\kappa]\frac{d}{dz}
  +2\kappa\omega\Lambda-\omega-E\right\}\varphi_+(z)=-\Delta^2 \varphi_+(z). 
\eeqa
This is a 4th-order differential equation of Fuchs' type. Explicitly,
\beqa
&&g^2z^2\frac{d^4\varphi_+}{dz^4}+4g^2\left[\frac{\omega}{g}(\Lambda-1) z^2+\left(q+\frac{1}{2}\right)z\right]\frac{d^3\varphi_+}{dz^3}\n
&&~~~~+\left\{4\omega^2(\Lambda^2-3\Lambda+1)z^2+4\omega g\left[3\left(\kappa+\frac{1}{2}\right)\Lambda-3\kappa-1\right]z
   +4g^2\kappa\left(\kappa+\frac{1}{2}\right)\right\}\frac{d^2\varphi_+}{dz^2}\n
&&~~~~+\left\{8\frac{\omega^3}{g}\Lambda(1-\Lambda)z^2+\left[8\omega^2\kappa(1-\Lambda)
   +8\omega^2\left(\kappa+\frac{1}{2}\right)(1-\Lambda)^2\right.\right.\n
&&~~~~~~~~~~\left.\left.     +4\omega(E-2\omega\kappa)\right]z  
   +8\omega g\kappa\left[\left(\kappa+\frac{1}{2}\right)\Lambda-\kappa\right]\right\}\frac{d\varphi_+}{dz}\n
&&~~~~+\left\{4\frac{\omega^2}{g}(1-\Lambda)\left(2\kappa\omega\Lambda-\omega-E\right)z\right.\n
&&~~~~~~~~~~\left. +4\omega^2\kappa^2(1-\Lambda)^2-\left[E-2\omega\left(\kappa-\frac{1}{2}\right)\right]^2+\Delta^2\right\}
   \varphi_+=0.\label{2-mode-diff1}
\eeqa
 As we show below, this equation
is quasi-exactly solvable provided that the system parameters $\Delta, \omega$ and $g$ satisfy
certain constraints, and exact solutions are polynomials in $z$. To this end,
we seek polynomial solutions of the the form to the differential equation (\ref{2-mode-diff1}), 
\beqa
\varphi_+(z)&=&\pd_{i=1}^{\cal M}(z-z_i),~~~~{\cal M}=1,2,\cdots,
\eeqa
where ${\cal M}$ is the degree of the polynomial and $z_i$ are roots of the polynomial to be determined. 
Following the procedure similar to that in the last section, we obtain the energies
\beqa
&&E=-\omega+\left[2{\cal M}+2\left(\kappa-\frac{1}{2}\right)+1\right]\omega\Lambda,\label{2-mode-energy}
\eeqa
and the constraint for the model parameters $\Delta, \omega$ and $g$, 
\beqa
\Delta^2+4\omega^2(1-\Lambda)\left[{\cal M}({\cal M}+2\kappa-1)+\frac{2\omega}{g}
   \Lambda\sum_{i=1}^{\cal M}z_i\right]=0.   \label{2-mode-constraint}
\eeqa
Here the roots $z_i$ satisfy the following system of algebraic equations,
\beqa
&&g^2z_i^2\sum_{p\neq l\neq j\neq i}^{\cal M}\frac{4}{(z_i-z_p)(z_i-z_l)(z_i-z_j)}\n
&&~~~~+4g\left[\omega(\Lambda-1) z_i^2+g\left(\kappa+\frac{1}{2}\right)z_i\right]
   \sum_{l\neq j\neq i}^{\cal M}\frac{3}{(z_i-z_l)(z_i-z_j)}\n
&&~~~~+\left\{4\omega^2(\Lambda^2-3\Lambda+1)z_i^2
   +4\omega g\left[3\left(\kappa+\frac{1}{2}\right)\Lambda-3\kappa-1\right]z_i
   +4g^2\kappa\left(\kappa+\frac{1}{2}\right)\right\}\sum_{j\neq i}^{\cal M}\frac{2}{z_i-z_j}\n
&&~~~~+8\frac{\omega^3}{g}\Lambda(1-\Lambda)z_i^2+8\omega^2\left[{\cal M}\Lambda
   +\left(\kappa+\frac{1}{2}\right)\Lambda(\Lambda-2)+\kappa\right]z_i\n  
&&~~~~~~~~~   +8\omega g\kappa\left[\left(\kappa+\frac{1}{2}\right)\Lambda-\kappa\right]~=~0,~~~~~~~~
  i=1,2,\cdots {\cal M}.\label{2-mode-BEs}
\eeqa
Here we have also used the relation (\ref{2-mode-energy}) in obtaining (\ref{2-mode-constraint}) 
and (\ref{2-mode-BEs}). The corresponding wavefunction component $\psi_+(z)$ is given by
\beq
\psi_+(z)=e^{-\frac{\omega}{g}  (1-\Lambda) z}\prod_{i=1}^{\cal M}(z-z_i)
\eeq
and the other component is $\psi_-(z)= e^{-\frac{\omega}{g}  (1-\Lambda) z}\varphi_-(z)$ 
with $\varphi_-(z)$ determined
from the first equation of (\ref{2-mode-diff}) for $\Delta\neq 0$,
$\varphi_-(z)=-\frac{1}{\Delta}{\cal L}_{2-m}\,\varphi_+(z)$. 
Because for any positive integer $n$, ${\cal L}_{2-m}z^n\sim z^n+{\rm lower~order~terms}$,
$\varphi_-(z)$ automatically belongs to the same invariant subspace as $\varphi_+(z)$.

As an example of the above general expressions, we consider the ${\cal M}=1$ case. The energy is
\beqa
&&E=-\omega+(2\kappa+2)\omega\Lambda.
\eeqa
(\ref{2-mode-BEs}) becomes
\beqa
&&\omega^2\Lambda(1-\Lambda)z_1^2+\omega g\left[\Lambda+\kappa
   +\left(\kappa+\frac{1}{2}\right)\Lambda(\Lambda-2)\right]z_i
   +g^2\kappa\left[\left(\kappa+\frac{1}{2}\right)\Lambda-\kappa\right]~=~0.\n
\eeqa
It has two solutions
\beq
z_1=-\frac{\kappa g}{\omega\Lambda},~~~~~~\frac{g\kappa(1-\Lambda)-g\Lambda}{\omega(1-\Lambda)}.
\eeq
Substituting into (\ref{2-mode-constraint}) gives the constraints $\Delta=0$ and
\beq
\Delta^2+8\kappa\omega^2=8\left(\kappa+\frac{1}{2}\right)\omega^2\Lambda^2=0,
    \label{2-mode-constraint-example}
\eeq
respectively. The constraint $\Delta=0$ corresponds to the case of degenerate atomic levels.
The constraint (\ref{2-mode-constraint-example}) is the non-trivial one. The corresponding
wavefunction $\psi_+(z)$ is given by
\beq
\psi_+(z)=e^{-\frac{\omega}{g}(1-\Lambda)z}
   \left(z-\frac{g\kappa(1-\Lambda)-g\Lambda}{\omega(1-\Lambda)}\right).
\eeq

\vskip.3in
\noindent {\bf Acknowledgments:} This work was supported by the Australian Research Council through 
Discovery Projects grant DP110103434.

\bebb{99}

\bbit{Vedral06}
V. Vedral, Modern foundations of quantum optics, Imperial College Press, London, 2006.

\bbit{Englund07}
D. Englund et al, Nature {\bf 450}, 857 (2007).

\bbit{Niemczyk10}
T. Niemczyk et al, Nature Phys. {\bf 6}, 772 (2010).

\bbit{Khitrova06}
G. Khitrova, H.M. Gibbs, M. Kira, S.W. Koch and A. Scherer, Nature Phys. {\bf 2}, 81 (2006).

\bbit{Leibfried03}
D. Leibfried, R. Blatt, C. Monroe and D. Wineland, Rev. Mod. Phys. {\bf 75}, 281 (2003).

\bbit{Braak11}
D. Braak, Phys. Rev. Lett. {\bf 107}, 100401 (2011).

\bbit{Solano11}
E. Solano, Physics {\bf 4}, 68 (2011).

\bbit{Moroz12}
A. Moroz, Europhys. Lett. {\bf 100}, 60010 (2012). 

\bbit{Maciejewski12}
A.J. Maciejewski, M. Przybylska and T. Stachowiak, 
arXiv:1210.1130v1 [math-ph];  
arXiv:1211.4639v1 [quant-ph].

\bbit{Travenec12}
I. Trav\'enec, Phys. Rev. A {\bf 85}, 043805 (2012).

\bbit{Chen12}
Q.H. Chen, C. Wang, S. He, T. Liu and K.L. Wang, Phys. Rev. A {\bf 86}, 023822 (2012).

\bbit{Moroz13}
A. Moroz, 
arXiv:1302.2565v2 [quant-ph], Ann. Phys., in press.

\bbit{Turbiner88}
A. Turbiner, Comm. Math. Phys. {\bf 118}, 467 (1988); 
arXiv:hep-th/9409068.

\bbit{Ushveridze94}
A.G. Ushveridze, Quasi-exactly solvable models in quantum mechanics, Institute of Physics
Publishing, Bristol, 1994.

\bbit{Gonzarez93}
A. Gonz\'arez-L\'opez, N. Kamran and P. Olver, Comm. Math. Phys. {\bf 153}, 117 (1993).

\bbit{Schweber67}
S. Scheweber, Ann. Phys. {\bf 41}, 205 (1967).

\bbit{Reik82}
H.G. Reik, H. Nusser and L.A. Ribeiro, J. Phys. A: Math. gen. {\bf 15}, 3431 (1982).

\bbit{Kus85}
M. Kus, J. Math. Phys. {\bf 26}, 2792 (1985).

\bbit{Ng99}
K.M. Ng, C.F. Lo and K.L. Liu, Eur Phys. J. D {\bf 6}, 119 (1999).

\bbit{Emary02}
C. Emary and R.F. Bishop, J. Math. Phys. {\bf 43}, 3916 (2002); J. Phys. A: Math. Gen. {\bf 35}, 8231 (2002).

\bbit{Sasaki07}
R. Sasaki, 
arXiv:0712.2616v1 [nlin.SI].

\bbit{Lee10}
Y.-H. Lee, W.-L. Yang and Y.-Z. Zhang, J. Phys. A: Math. Theor. {\bf 43}, 185204 (2010); 
ibid {\bf 43}, 375211 (2010).

\bbit{Lee11}
Y.-H. Lee, J.R. Links and Y.-Z. Zhang, Nonlinearity {\bf 24}, 1975 (2011).

\bbit{Zhang12}
Y.-Z. Zhang, J. Phys. A: Math. Theor. {\bf 45}, 065206 (2012).

\bbit{Zhang13a}
Y.-Z. Zhang, 
arXiv:1304.3979v2 [quant-ph], J. Phys. A: Math. Theor., in press.

\bbit{Gerry85}
C. C. Gerry, Phys. Rev. A {\bf 31}, 2721 (1985); ibid {\bf 37}, 2683 (1998).

\bbit{Dolya09}
S.N. Dolya, J. Math. Phys. {\bf 50}, 033512 (2009).

\eebb

\end{document}